# Mapping Dynamical Magnetic Responses of Ultra-thin Micron-size Superconducting Films using Nitrogen-vacancy Centers in Diamond

*Ying Xu[1], Yijun Yu[1], Yuen Yung Hui[2], Yudan Su[1], Jun Cheng[1], Huan-Cheng Chang[2], Yuanbo Zhang[1,3], Y. Ron Shen[1,4], Chuanshan Tian*[1,3]*

[1] Department of Physics, State Key Laboratory of Surface Physics and Key Laboratory of Micro- and Nano-Photonic Structures (MOE), Fudan University, Shanghai 200433, China

[2] Institute of Atomic and Molecular Sciences, Academia Sinica, Taipei

[3] Collaborative Innovation Center of Advanced Microstructures, Nanjing, 210093, China

[4] Department of Physics, University of California, Berkeley, California 94720, United States

**Abstract:** Two-dimensional superconductors have attracted growing interest because of their scientific novelty, structural tunability, and useful properties. Studies of their magnetic responses, however, are often hampered by difficulties to grow large-size samples of high quality and uniformity. We report here an imaging method that employed NV⁻ centers in diamond as sensor capable of mapping out the microwave magnetic field distribution on an ultrathin superconducting film of micron size. Measurements on a 33nm-thick film and a 125nm-thick bulk-like film of $Bi_2Sr_2CaCu_2O_{8+\delta}$ revealed that the ac Meissner effect (or repulsion of ac magnetic field) set in at 78K and 91K, respectively; the latter was the superconducting transition temperature ($T_c$) of both films. The unusual ac magnetic response of the thin film presumably was due to thermally excited vortex-antivortex diffusive motion in the film. Spatial resolution of our ac magnetometer was limited by optical diffraction and the noise level was at 14 $\mu T/Hz^{1/2}$. The technique could be extended with better detection sensitivity to extract local ac conductivity/susceptibility of ultrathin or monolayer superconducting samples as well as ac magnetic responses of other two-dimensional exotic thin films of limited lateral size.

**Introduction**

Repelling of magnetic flux (the Meissner effect) is usually taken as the concrete evidence of superconductivity present in a material. In the ac case, study of flux repulsion or magnetic field distribution also allows extraction of ac susceptibility/conductivity of superconducting films. The latter are important as they also provide crucial information about the superfluid density, vortex/antivortex excitations, and fluctuation-induced effects in the films.[1-6] Recently, thin-film and two-dimensional (2D) superconductors (SC) have attracted tremendous interest,[7-12] but often because the sample size is small, study of magnetic response is very challenging. Common schemes for such measurements involving the use of two-coil mutual inductance[1,2] and microwave (MW) cavity,[13,14] etc, require good uniform samples of millimeter(mm) or larger lateral size; recently discovered 2D or quasi-2D superconductors, however, are often limited to tens of microns.[9,12] A technique that can map local ac magnetic response of SC on a few μm scale is in urgent need.

We have developed a technique based on the high sensitivity of negatively charged nitrogen vacancy (NV⁻) centers in diamond to detect magnetic field. It allows us to map out the distribution of ac magnetic field influenced by a thin-film SC with a spatial resolution 0.5 μm. The NV⁻ centers in diamond have emerged in recent years as an extremely sensitive magnetic field sensor with spatial resolution on the atomic scale if single NV⁻ centers are used and on the sub-micron scale limited by optical diffraction if groups of NV⁻ centers are used.[15-17] The technique has already been adopted in a number of pioneering works to probe the dc Meissner effect and image the vortex lines in mm-size SC films.[18-21] Study of μm-size samples and their ac magnetic responses however has not yet been explored.

We report here a successful application of the technique to an exfoliated $Bi_2Sr_2CaCu_2O_{8+\delta}$(BSCCO) flake of ~10μm lateral size consisting of several regions with thickness varying from 25nm to 33nm.[9] (The thickness of 1 unit cell is ~3nm.) The sample was situated in a microwave (MW) field with its magnetic component perpendicular to the surface and the sample temperature could be varied from 300K to 40K. The magnetic field distribution immediately above the surface of the flake at different temperatures was measured with a spatial resolution of ~0.5μm and a detection sensitivity of 14μT/Hz$^{1/2}$. The results agreed reasonably well with an approximate calculation below the superconducting transition temperature ($T_c$). However, unlike in a bulk-like film, the magnetic flux repulsion in the thin film did not occur until 13K below $T_c$. This characteristic behavior of thin SC films in an ac field has been reported in the literature, and is believed to be due to diffusive motion of thermally excited vortices and anti-vortices in the films. Our technique already has a sensitivity to probe ac magnetic responses of monolayer SC. It can be further improved to apply to studies of ac conductivity/susceptibility, vortex dynamics and fluctuation-induced effects in thin film SC.

**Underlying Principle of the Technique and Experimental Arrangement**

We follow the general scheme of optical detection of magnetic resonance by NV⁻ centers in diamond. The relevant energy states of the NV⁻ centers are depicted in Fig. 1(a). In the absence of static magnetic field, the ground state manifold is composed of the spin singlet $m_s= 0$ ($|g,0\rangle$) and a degenerate doublet $m_s= \pm 1$ ($|g,\pm 1\rangle$) 2.88GHz above $|0\rangle$ at cryogenic temperature.[22] Populations in $|g,0\rangle$ and $|g,\pm 1\rangle$ can be measured through detection of photoluminescence (PL): A green laser beam (532 nm) excites electrons to excited electronic states and spins are conserved in the excitation and ensuing luminescence processes. Electrons excited from $|g,0\rangle$ first end up by relaxation mainly in the first excited state $|e,0\rangle$, followed by luminescence to $|g,0\rangle$. Electrons excited from $|g,\pm 1\rangle$ also relaxes to $|e,\pm 1\rangle$ followed by luminescence to $|g,\pm 1\rangle$, but a significant part of them goes through nonradiative relaxation first to an intermediate singlet state $|s\rangle$ and ends up in $|g,0\rangle$. The nonradiative leakage reduces the luminescence efficiency from $|e,\pm 1\rangle$ by ~30%. The difference in luminescence efficiencies from excitations of $|g,0\rangle$ and $|g,\pm 1\rangle$ can therefore be used to obtain the populations in $|g,0\rangle$ and $|g,\pm 1\rangle$.

Initially, if a long excitation pulse is applied to the NV⁻ system, the population will all be accumulated in |g,0⟩ through optical pumping. Subsequently, switching on a ~2.88 GHz MW field will induce transient nutation that appears as population oscillation between |g,0⟩ and |g,±1⟩ at the Rabi frequency, which depends on the MW field strength.[17,23] Thus, PL detection of the population oscillation allows us to deduce the magnetic field of the MW.

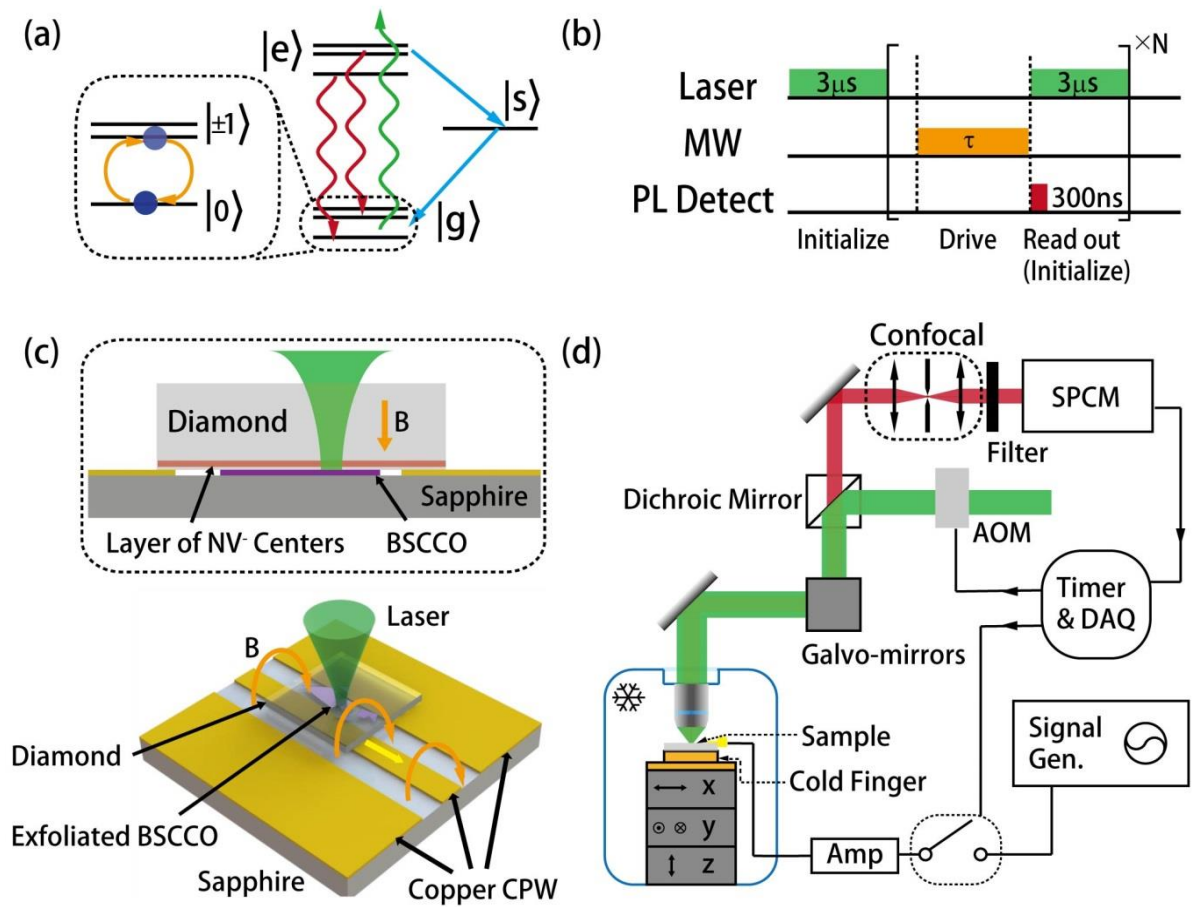

**Figure 1**: (a) Simplified energy level diagram of a NV⁻ center. Green and red wavy lines denote excitation at 532 nm and broadband luminescence from ~630nm to ~800nm, respectively, and the solid blue line describes nonradiative relaxation. Spin sublevels, $|0\rangle$ and $|\pm1\rangle$, of the ground manifold $|g\rangle$ are expanded in the dashed box. Yellow arrows indicate the coherent population oscillation between spin states induced by MW. (b) Pulse sequence used to measure the Rabi oscillation of populations between $|0\rangle$ and $|\pm1\rangle$. (c) Sample and beam geometry. The BSSCO flake on the NV-diamond chip was prepared by exfoliation. The yellow arrow represents high frequency current flowing in the central conductor. The MW magnetic field direction is illustrated by orange arrow. The 532nm laser beam is normally incident. (d) Schematic of the experimental setup. Laser was gated by acousto-optical modulator (AOM), and the MW pulses were delivered through coaxial cable which is connected to the CPW. PL signal is collected by single photon counting module (SPCM).

Experimentally, we used the pulse sequence in Fig.1(b) for PL measurements. The 3 μs green laser pulse was first applied to optically pump the NV- centers to $|g,0\rangle$, followed by a MW pulse of variable duration τ to induce Rabi population oscillation between $|g,0\rangle$ and $|+\rangle=(|g,+1\rangle+|g,-1\rangle)/\sqrt{2}$. At the end of the MW pulse, the 3 μs laser pulse was applied again to induce PL and to prepare for the next cycle of measurement. The PL was collected at the leading 300 ns of the laser pulse, during which population relaxation of $|g,0\rangle$ and $|g,\pm1\rangle$ was negligible. To observe the Rabi oscillation, we scanned τ from 0 to 2 μs. At each τ, we repeated the PL measurement $3\times10^5$ times to achieve a good signal-to- noise ratio.

Figure 1(c) and 1(d) illustrate the sample and beam geometry and the experimental arrangement. The (100) CVD-grown diamond plate (2mm×2mm×0.3mm) was implanted with $^{15}N_2^+$ ions at 40keV with a dosage of $4\times10^{14}/cm^2$, and then annealed at 850°C in vacuum, resulting in a distribution of NV⁻ centers in a thin layer ~50nm below the surface. The thin BSCCO film on the diamond plate was prepared by exfoliation of a thicker film sample initially deposited on the diamond plate (see SI). The BSCCO/diamond sample was situated on top of the copper planar waveguide (CPW) at a place where the magnetic field of the MW in the waveguide was perpendicular to the diamond plate. MW pulses were delivered to the sample area through the CPW, which was deposited on a sapphire substrate. For low temperature measurement, the sapphire plate was clamped to the cold finger of a cryostat, with piezo xyz-positioner for position control, as shown in Fig.1(d). The laser beam was focused to a spot of ~5μm in diameter on the sample surface. The PL signal was collected through an objective lens (NA = 0.5) and detected by a single photon counting module (SPCM) after passing through a 630nm long-pass filter and a confocal pinhole. The spatial resolution of our system was

~430nm for 20%-80% edge response at PL wavelength ~650nm.[24] Mapping of the sample was achieved by scanning the laser spot with two galvo-mirrors.

**Results and Discussion**

Fig. 2(a) displays the optical image of an exfoliated BSCCO film on diamond. The film thickness was measured by atomic force microscopy (AFM) (Fig. 2(b)). It showed variation from 7nm to 33nm (Fig.2(c)) in different regions over the film, corresponding to a change of 2.5 to 11 unit cells of BSSCO. For transport measurement, four-probe electrodes were fabricated on the SC film. As seen in Fig. 2(d), the electric resistivity of our BSSCO film versus temperature exhibited a superconducting transition temperature at ~91K, essentially the same as that of bulk BSCCO.[25]

We first carried out a measurement with the BSSCO film in the normal state at 100K to find the MW magnetic field $B_0$ sensed by the NV$^-$ centers as a reference. The observed Rabi oscillation is shown in Fig. 2(e), from which we found $B_0$ = 170±2μT by fitting. To map out the field distribution on top of the BSSCO film in the SC state, we fixed the MW input with its pulse duration $\tau$ set at 75 ns, close to the maximum slope of the Rabi oscillation in Fig.2(e). This was to ensure maximum detection of $B$ field variation upon laser scanning over the SC film. It can be shown that the normalized difference PL signal, defined as $C=[I(\tau)-I(0)]/I(0)$ with $\tau$ = 75ns is proportional to $B^2$ when $B$ is smaller than $B_0$ (see SI). Here, $I(\tau)$ is the PL signal detected immediately after the MW pulse of duration $\tau$ is over and $I(0)$ refers to the PL signal in the absence of MW.

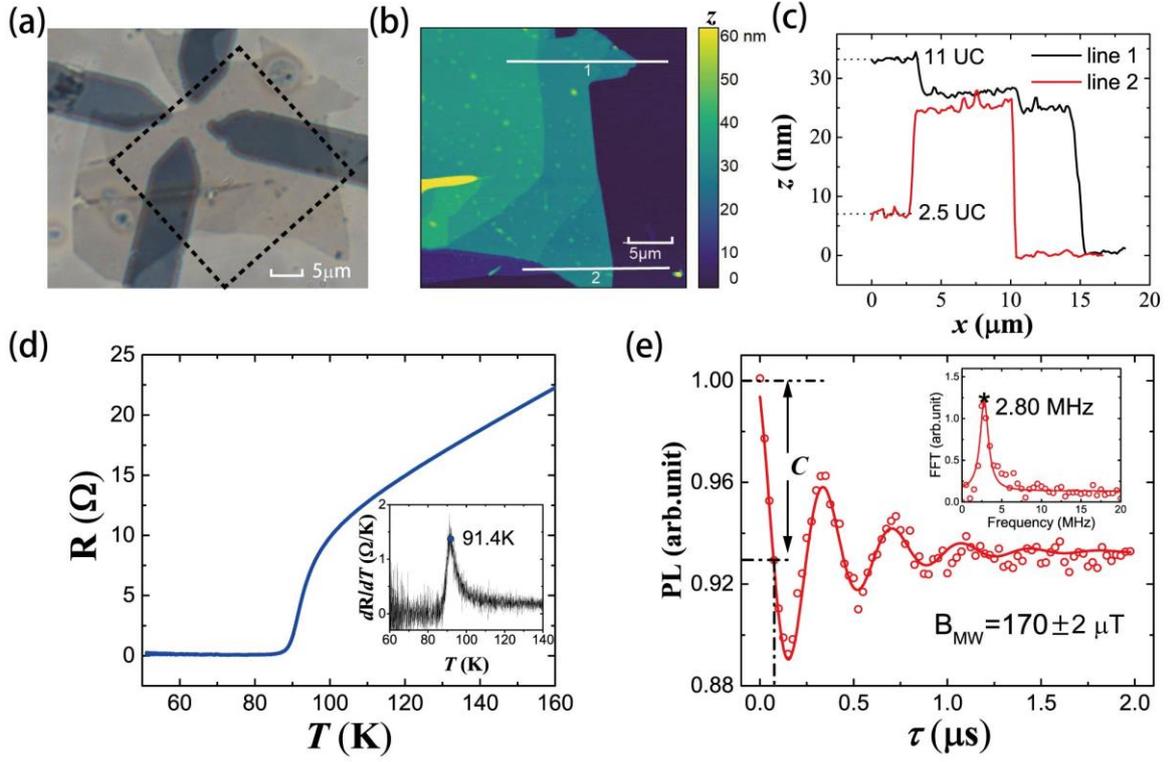

**Figure 2**: (a) optical image of a BSCCO flake. Dark gray cross is the gold electrodes used for transport measurement. Dashed box indicates the region where magnetic field mapping was carried on. (b) AFM image of the right side of the flake. Bright yellow stripe on the bottom left corner is a crack on the flake also observed in (a). (c) AFM profile along line 1 and line 2 in (b). (d) Resistance of the flake versus temperature showing the SC transition; derivatives of the curve shown in the inset yields a transition temperature at $T_c \sim$ 91K. (e) Rabi oscillation

measured with the flake in the normal state at 100K; inset: Fourier transformation of the Rabi oscillation curve.

As a reference, we first recorded the PL image of the BSSCO film at 80K without application of the MW pulse, presented in Fig. 3(a). The image resulted from local field changes seen by NV$^-$ centers due to variation of BSSCO film thickness. A line profile of the image described in Fig.3(b) shows the edge of the SC film, from which the resolution of our confocal optical system could be estimated to be 530 nm, close to the diffraction limit of the system. We then mapped out the MW magnetic field distribution on the BSSCO film at various temperatures, shown in Fig.3(c)-3(g) for $T$=90, 80, 75, 70 and 40K, respectively. The field distribution remained unchanged for $T>T_c$(=91K). Even down to 80K, no detectable change was observed, contrary to what occurred in bulk BSCCO, where magnetic field screening appeared immediately below $T_c$. As will be discussed later, this is attributed to vortex-antivortex diffusive motion present in a thin SC film.[3,6] Upon cooling below 80K, the magnetic field distribution on the BSSCO film changed rapidly. In the interior region of the film, reduction of the field strength was appreciable due to screening by supercurrent in the film, but the spatial variation was smooth. Near the film boundary and cracks, the field was significantly enhanced together with a large spatial gradient. The interior field decreased further at lower temperatures, but less appreciable after the temperature dropped below ~70K, and at 40K, reduced to nearly half of the field seen at 90K. No clear boundary was observed between domains of different thicknesses in all cases because smearing out of the magnetic flux obscured the difference across the boundary as will be explained below.

To provide an understanding of the observed results, we used the two-fluid model[13,26] and carried out a calculation of the field distribution on a thin SC disk of radius $r = 10\mu m$ and thickness $d = 30nm$. The ac conductivity, $\sigma(\omega,T) = \sigma_1 - i\sigma_2$, of SC is given by[26]

$$\sigma(\omega,T) = \sigma_1 - i\sigma_2$$
$$\sigma_1 = \frac{e^2\tau(T)}{m^*}n_n(T) = \frac{\tau_r(T)}{\mu_0\lambda^2(0K)}\left(1-\frac{\lambda^2(0K)}{\lambda^2(T)}\right) \quad (1)$$
$$\sigma_2 = \frac{e^2}{m^*\omega}n_s(T) = \frac{1}{\mu_0\lambda^2(0K)\omega}\frac{\lambda^2(0K)}{\lambda^2(T)}$$

Here, $n_n$ and $n_s$ are the density of normal and superconducting carriers respectively, $\tau_r$ is the relaxation time of the normal carriers (of the order of picosecond[13,26]), $m^*$ and $e$ are effective mass and charge of the carriers, and $\lambda$ is the London penetration depth. We have neglected the contribution of normal carriers to $\sigma_2$. The temperature dependence of $\tau_r$ and $\lambda^2/\lambda(0K)^2$ for bulk BSCCO is known from the literature.[26] The ac conductivity can be considered as pure imaginary when $\omega\tau_r$ ($10^{-2} \sim 10^{-3}$ at $T < T_c$ in our case) is much smaller than $\lambda^2/\lambda(0K)^2$, which is true for temperature 10K below $T_c$. The current induced by external MW field in the film is solely the diamagnetic current $\mathbf{J} = \mathbf{A}d/\mu_0\lambda^2$, where $\mathbf{J}$ is the sheet current density, $d$ is the film thickness and **A** is the vector potential; the field distribution is the result of ac Meissner screening.[27,28] Following the method developed by Brandt,[27,28] we calculated the MW magnetic field distribution at 70K and 40K, where the London penetration depth $\lambda$(0K) was taken to be 210nm (see details in SI).[26] The calculated radial field distribution is presented in Fig.3(h), together with the line-cuts of the observed field distribution image in Fig. 3(f) and 3(g). Qualitatively, both the calculation and the experiment show that the field is better screened in the interior of an SC thin film and the screening effect is stronger at lower temperature. At the edge of the film, the magnetic field was naturally enhanced. The calculation revealed a field penetration of ~60% of $B_0$ at 5 μm from the edge into the SC film at 40K. Because the change of the magnetic field across the boundary between two domains of different thicknesses is quite small and expected to be smooth, the boundary line was not observable in the magnetic field images of Fig.3(c)-(g).

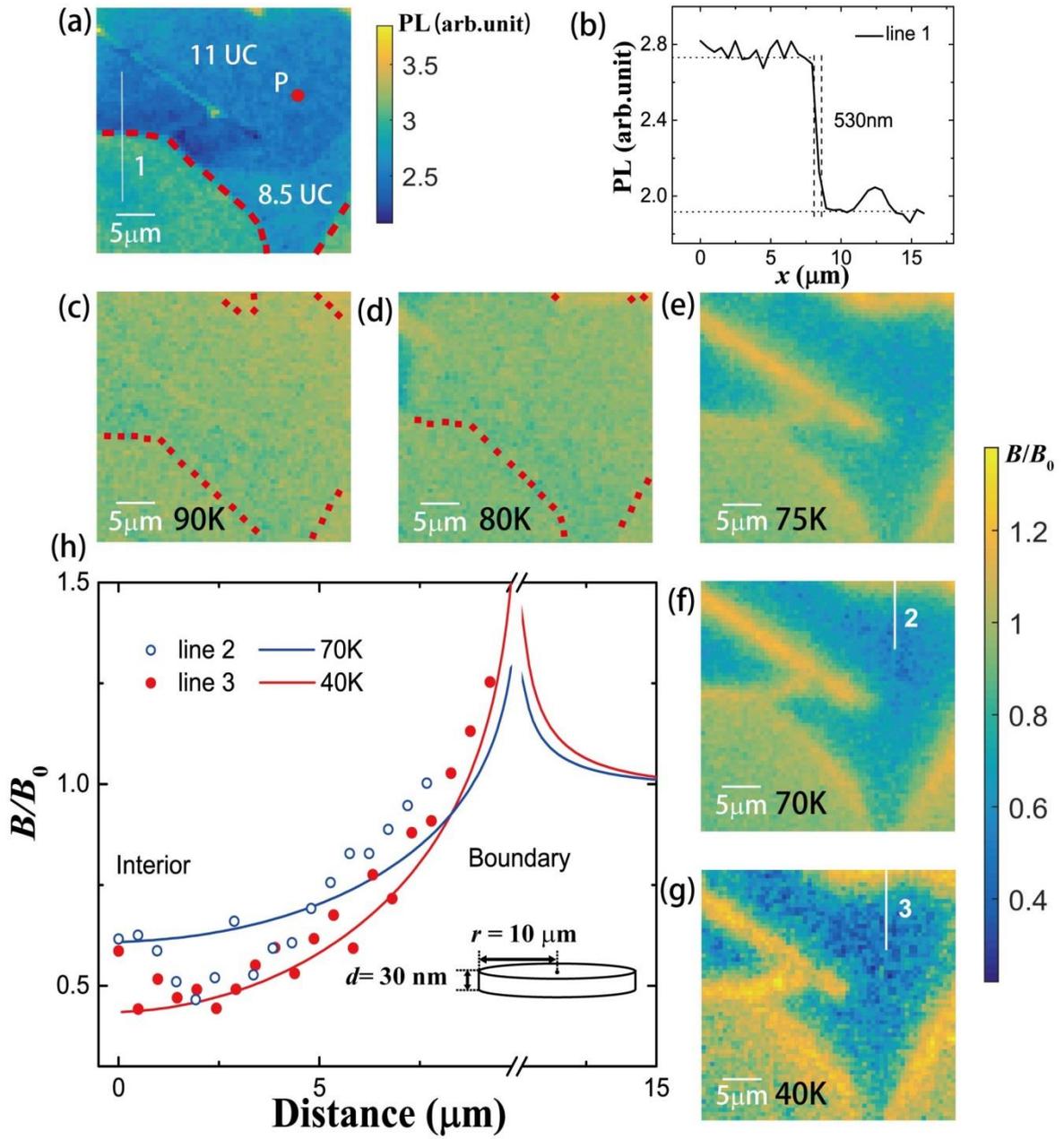

**Figure 3**: (a) Photo-luminescence (PL) image of an area (circled out in the dashed box of Fig.2(a)) of the BSSCO film at 80K. The dark blue region is the SC sample, and red dashed line indicates the edge of SC film. Red dot P indicates the point where we measured change of the local magnetic field versus temperature. (b) Profile of PL along the line across an edge of the film in (a). (c)-(g) Magnetic field images at different temperatures. Red dash lines in (c) and (d) indicate the edges of the film. h) Calculated magnetic field distribution on top of a SC circular disk with radius 10μm, thickness 30nm (sketched in the inset) at 40 K (red curve) and 70 K (blue curve). The magnetic field variations along line 2 in (f) and line 3 in (g) are shown for comparison with calculation.

We now compare the temperature dependence of magnetic field screening in the interior of the BSSCO thin film described above and a 125-nm BSSCO film that resembles a bulk. The measurement of the 125-nm film is described in the SI. As seen in Fig.4(a), screening sets in at $T^* = $ 78K and 91K in the two cases, respectively, and a theoretical curve calculated from the two-fluid model with the London penetration depth λ(0K) taken as 230 nm and λ(T) from the literature,[26] is shown for comparison. The experimental result of the bulk film agrees fairly well with the theory, but that of the thin film clearly deviates from the theory. Field screening of the thin film is only partial, but most importantly, it does not set in until ~13K below the SC transition temperature. The thin-film experimental result is compared with the theoretical calculations from the two-fluid model in Fig.4(b), where the measured resistance change with temperature for the thin film is also presented as a reference. The theory agrees with experiment below ~72K, but obviously disagrees in the range between 72 and 91K ($T_c$). Such discrepancy was commonly observed in measurements of ac magnetic responses of ultra-thin SC films.[3,29,30] It is generally understood to be a 2D effect due to diffusive motion of thermally excited vortex-antivortex.[3,6] Presumably because of reduced pinning or viscosity, the vortex-antivortex excitations are more diffusive for thinner SC films.[3,6] The diffusion reduces the supercurrent induced by the applied magnetic field.[3-6] Accordingly, the diamagnetic response of the SC thin film is greatly suppressed until the temperature is sufficiently low such that the vortex-antivortex diffusive motion becomes insignificant.[1,3,6]

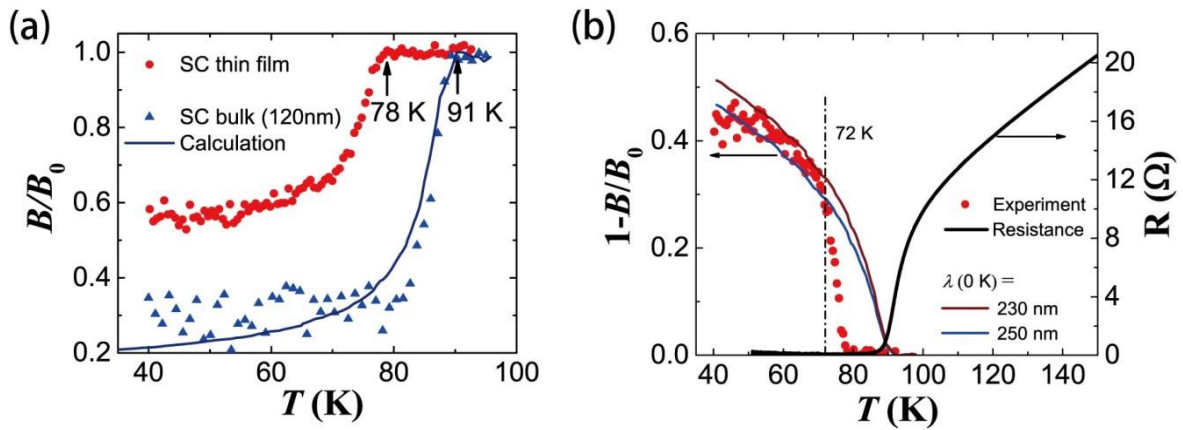

**Figure 4**: (a) Comparison of the measured local magnetic fields on the thin film (red dots, described in the text) and a 125-nm bulk film (blue dots) of BSSCO with a theoretical curve (blue) calculated from the two-fluid model. Black arrows mark the onsets of transition. (b) Comparison of the measured (red dots) and calculated magnetic field repelled by the thin film versus temperature. The calculation was based on the two-fluid model, and results with $\lambda$

assumed to be 230 and 250 nm are descried by the brown and blue curves, respectively. Resistance of the thin film versus temperature is shown for reference.

The technique described here actually has the sensitivity to probe magnetic response of a monolayer SC sample. As shown in Fig 4(a), the detection limit of our magnetic field measurement was about 0.01 $H_0$, i.e. 1.7μT, when each data point was obtained by averaging over repeated measurements in 72 sec, corresponding to a noise level of 14μT/Hz$^{1/2}$. Such detection sensitivity is capable of mapping MW magnetic field distribution on a monolayer SC (~1.5nm, half unit cell) with lateral dimension of ~10μ. Based on the theoretical model given above, the magnetic field repelled is roughly proportional to the ratio of the lateral dimension *a* and the Pearl penetration depth $\Lambda = 2\lambda^2/d$,[6,27,28] when $a < \Lambda$ (see SI), and therefore is proportional to the film thickness *d*. For a monolayer thin film with the same lateral dimension as the thin film studied here, the field repelled would be only ~ 1/20 of what we measured, but could still be well resolved with ~100s data averaging. At present, we were not able to observe field screening from the 7nm-thick flake shown in Fig. 2(b). This was because the flake was too small in lateral dimension (~4μm in width) and rapidly degraded in air.[9]

Our technique can be further significantly improved. Much better detection sensitivity can be realized if the coherence time of Rabi oscillation of NV$^-$ centers can be longer to allow recording of more cycles of oscillation. Currently, the coherence time of our NV⁻ centers was limited to ~1μs, which is 10 times less than the reported value on high-quality NV⁻ diamond samples.[15,17] The sensitivity can also be drastically enhanced using dynamical decoupling pulse sequence schemes.[15] On characterization of SC films, sensing of local MW magnetic response of the films allows deduction of local ac conductivity/susceptibility of the films if both amplitude and phase of the magnetic field can be measured.[27,28] Such magnetic field sensing by NV⁻ magnetometry has already been demonstrated.[17] Finally, the technique is not limited to a fixed MW frequency. The spin states of the ground level of NV⁻ can be split and tuned by an applied static magnetic field with a *g*-factor of ~28GHz/T, and the transition frequency between the spin states used to probe the ac response through Rabi oscillation varies accordingly. With the dc magnetic field changed from 0 to 0.3 T, the frequency can vary from 10 MHz to 10 GHz. Note that possible application to SC films relies on the assumption that the dc magnetic field would not affect the ac properties of the films. Magnetic response in the radio-frequency range (~10kHz to ~1MHz) is also possible with the spin-echo scheme, where the frequency can be tuned continuously without the introduction of a static magnetic field.[15]

**Conclusion**

In summary, we have demonstrated the use of NV- diamond as a microscopic sensor to map out MW magnetic field distribution on a thin-film, quasi-2D SC of micron size in lateral dimension and observe the ac Meissner effect. The device has a spatial resolution limited by optical diffraction and a sensitivity capable of resolving the magnetic field redistribution influenced by a 10-μm monolayer SC film. Further improvement of the technique to enable it to detect direction, amplitude and phase of the local magnetic field will allow extraction of the local ac conductivity/susceptibility of SC films. This magnetometer could become a most powerful tool to investigate ac magnetic responses, including vortex dynamics and phase fluctuations of complex order parameter, of novel 2D or quasi-2D superconductors that are only available with limited sample size, e.g., bi-layer and tri-layer graphene, $MoS_2$, $TaSe_2$, etc. More generally, it can also be used to probe ac magnetic properties and conductivity/susceptibility of non-SC 2D/quasi-2D micro-scale materials.